\documentclass{aa}  

\usepackage[utf8]{inputenc}
\usepackage{amssymb,amsmath}
\usepackage{graphicx}
\usepackage[dvipsnames]{xcolor}
\usepackage[breaklinks=true]{hyperref}
\hypersetup{colorlinks=true,allcolors=teal}
\usepackage[flushleft]{threeparttable}
\usepackage{footmisc}
\usepackage{subfigure}
\usepackage{multirow}

\newcommand{\be}{\begin{equation}}
\newcommand{\ee}{\end{equation}}
\def\bear#1\ear{\begin{align}#1\end{align}}
\newcommand{\nline}{\notag \\}
\newcommand{\f}{\frac}
\newcommand{\de}{\mathrm{d}}

\usepackage[T1]{fontenc}
\usepackage{amsmath}	
\DeclareRobustCommand{\VAN}[3]{#2}
\let\VANthebibliography\thebibliography
\def\thebibliography{\DeclareRobustCommand{\VAN}[3]{##3}\VANthebibliography}
\usepackage{graphicx}
\usepackage{txfonts}
\begin{document}

   \title{Prospects of FRBs and Large Scale 21~cm Power Spectra in Constraining the Epoch of Reionization}


   \author{Barun Maity
          \inst{1}}

   \institute{Max-Planck-Institut f\"ur Astronomie, K\"onigstuhl 17, D-69117 Heidelberg, Germany\\
              \email{maity@mpia.de}}
              

   \date{Received XXX; accepted XXX}

 
  \abstract{The Epoch of Reionization (EoR) is a crucial link to grasp the complete evolutionary history of the universe. Several attempts with a variety of observables have been utilized in the past to understand the thermal and ionization evolution of the Intergalactic Medium during EoR. In this study, we explore the simultaneous prospects of two important observables which are expected to be available in the near future, i.e. Dispersion Measure (DM) of high redshift FRBs and large scale 21 cm power spectra. For this purpose, we use an earlier developed explicitly photon conserving semi numerical model, \texttt{SCRIPT} including realistic recombination and radiative feedback effect. We check that DM evolution of 100 mock FRBs at high redshifts ($7.0\le z\le15.0$) is sufficient to recover the underlying reionization model, while 1000 FRB mocks at redshift range can constrain the reionization timeline within the percentage level uncertainties at 68\% confidence limit. Further, we study the effect of including large scale 21~cm power spectra (using only a single bin, $k\sim0.14~h/\mathrm{cMpc}$) at three redshifts along with FRB DM distribution. The joint exploration using these two observables can significantly improve the constraints on the various parameters ($\lesssim 8\%$ uncertainties for reionization interval and midpoint at 95\% confidence) alleviating the degeneracies and can narrow down the thermal history of the universe by discarding some of the extreme heating models.}
   

   \keywords{intergalactic medium -- cosmology: theory – dark ages, reionization, first stars -- large-scale structure of Universe}

   \maketitle
%

\section{Introduction}
\label{sec:intro}
The Epoch of Reionization (EoR) marks a pivotal phase in the cosmic timeline, bridging the early universe characterized by predominantly linear structures with the later stages dominated by intricate and non-linear astrophysical phenomena. During this epoch, the universe undergoes a transition from a largely neutral to a mostly ionized state, driven by the emission of ionizing photons from the earliest luminous sources \citep[for reviews, see][]{2001PhR...349..125B,2009CSci...97..841C,2018PhR...780....1D,2022arXiv220802260G,2022GReGr..54..102C}.  Despite significant advancements in the field, the precise timeline of reionization and the nature of the primary sources responsible for this transformative process remain elusive, posing compelling questions.

There already exists a wide variety of observational probes spanning across multiple wavelengths to explore the reionization epoch \citep{2013ASSL..396...45Z}. In the near future, Fast Radio Bursts (FRBs) can be an important probe of the epoch of reionization complementing the others. These are extremely bright radio transients \citep{2007Sci...318..777L} coming from different parts of the universe. Although the progenitors of these extreme sources are yet to be fully understood, these have been detected at a wide range of frequencies allowing us to trace the Intergalactic medium (IGM) at different distances \citep{2020Natur.581..391M}. The distribution of the high redshift FRBs is expected to provide useful information regarding the ionization state of the IGM. This is motivated by the fact that the maximum contribution to the Dispersion Measure (DM) of a high redshift FRB is essentially the line of sight integration over the ionization history of the IGM. Hence, in recent times, detecting the high redshift FRBs is one of the major science goals of the current and upcoming observational facilities such as the Canadian Hydrogen Intensity Mapping Experiment \citep[CHIME;][]{2018ApJ...863...48C}, the Five-hundred-meter Aperture Spherical radio Telescope \citep[FAST;][]{2019RAA....19...16L}, the Australian Square Kilometre Array Pathfinder \citep[ASKAP;][]{2009ASPC..407..446J}, the transient Universe in real Time with Molongo
Observatory Synthesis Telescope \citep[UTMOST;][]{2017PASA...34...45B}, MeerKAT \citep{2009arXiv0910.2935B}, Square Kilometer Array \citep[SKA;][]{2015aska.confE..55M}. So far, FRBs have been observationally detected up to redshift, $z<2$ \citep{2018MNRAS.475.1427B,2023Sci...382..294R}. However, with the advancement of future observational facilities, we can expect to have FRBs at higher redshifts which can probe EoR \citep{2018ApJ...867L..21Z,2020MNRAS.494..665L}.

On the other hand, the 21 cm signal arising from the fluctuation in the neutral hydrogen (HI) field has a strong potential to provide the nature of the IGM and ionizing sources during the reionization era.  The radio interferometric observations have already been able to put stringent limits on the amplitude of the large scale HI fluctuations. These observations include Low Frequency Array \citep[LOFAR;][]{2019MNRAS.488.4271G,2020MNRAS.493.1662M}, Murchison Widefield Array \citep[MWA;][]{2019ApJ...884....1B,2020MNRAS.493.4711T}, Giant Metrewave Radio Telescope \citep[GMRT;][]{2013MNRAS.433..639P} and Hydrogen Epoch of Reionization Array \citep[HERA phase I;][]{2022ApJ...924...51A}. In the near future, telescopes like the Square Kilometer Array \citep[SKA-Low;][]{2015aska.confE...1K} and the fully deployed HERA \citep{2017PASP..129d5001D} will aim for the direct mapping of the signal. Also, there exist experiments like Owens Valley Long Wavelength Array \citep[OVRO-LWA][]{2019AJ....158...84E}  and New Extension in Nançay Upgrading LOFAR \citep[NenuFAR][]{2021sf2a.conf..211M} aiming for the detection of 21~cm signal from higher redshift epoch i.e. Cosmic Dawn.

Hence, it is necessary to have realistic reionization models to properly interpret the observational data. There are several ways to model the Epoch of Reionization starting from simplistic and very fast analytical models to accurate and computationally expensive full radiative transfer simulation. The analytical models dealing with global evolution of the ionization fraction have been widely used to pursue forecasting on hydrogen reionization with FRBs \citep{2016JCAP...05..004F,2021MNRAS.502.5134B,2021JCAP...05..050D,2021MNRAS.502.2346H,2021ApJ...906...49Z,2022ApJ...933...57H}. However, as these models deal only with the globally averaged ionization state, they are not able to track the fluctuations in the IGM due to reionization lacking important information about the nature of the IGM. On the other side, the rigorous numerical simulations are not efficient enough to pursue full parameter space explorations with multiple observables. At the middle of these two extremes,  an efficient way to model the reionization epoch containing the important inhomogeneous effects for different astrophysical processes is to utilize semi numerical approaches. As reionization is intrinsically inhomogeneous in nature, these effects can be critical to correctly glean information from the observational data. The main idea of these semi numerical models is to implement a photon counting algorithm given the source information and extract the ionization field by comparing with neutral hydrogen distribution \citep{2007ApJ...669..663M,2008MNRAS.386.1683G,2010MNRAS.406.2421S,2011MNRAS.411..955M,2013ApJ...776...81B}.  Earlier studies utilizing semi numerical models have shown the promises of FRBs to rule out some extreme reionization topologies \citep{2021MNRAS.505.2195P}. These models also allow us to exploit 21~cm power spectra as an additional probe along with FRB DM distribution. However, most of the semi numerical models adopt excursion set based techniques for efficient photon counting which are known to face the problem of photon number non conservation \citep{2007ApJ...654...12Z, 2011MNRAS.414..727Z,2016MNRAS.460.1801P}. Recently, a novel way to tackle this problem has been introduced in a recent study with an explicit photon conserving approach \citep{2018MNRAS.481.3821C}.  

In this work, we utilize the reionization model \texttt{SCRIPT} (\textbf{S}emi Numerical \textbf{C}ode for \textbf{R}e\textbf{I}onization with \textbf{P}ho\textbf{T}on Conservation), based on photon conserving algorithm which also captures important physical effects like recombination and photoionization feedback \citep{2022MNRAS.511.2239M}. The model has been shown to be extremely promising in inferring  the reionization physics using a variety of observables \citep{2022MNRAS.515..617M}. The photon conserving model is also capable of extracting the astrophysical parameters and state of the IGM using 21~cm mock data even with a relatively coarse resolution simulation \citep{2023MNRAS.521.4140M}. Here, we check the prospects of highly dispersed FRBs originated at EoR redshifts in constraining both the ionization and thermal state of the IGM during reionization. Specifically, we generate DM  mocks with 100, 500, and 1000 FRBs and study the comparison of constraining power. Further, we include the information of large scale 21~cm power spectra ($k\sim0.14~h/\mathrm{cMpc}$) to pursue a joint forecast with FRB DM. From the status of the current and upcoming interferometer experiments (HERA, SKA-low, etc), this fiducial choice is a realistic one where we can expect the detection of a 21 cm signal. This study provides us an realistic expectations from the upcoming observational facilities in understanding the epoch of reionization.

This paper is organized as follows: In Section \ref{sec:theory},  we describe the basic theoretical framework and provide a brief overview of the \texttt{SCRIPT} model parameters. In Section \ref{sec:DM_FRB}, we lay out the procedure of computing dispersion measures of FRBs using our reionization models followed by a discussion on the variation of dispersion measure with respect to different parameters in Section \ref{sec:DM_variation}. Then, in Section \ref{sec:generate_mock}, we describe the methodology used to generate the FRB DM and 21~cm mock data. This is followed by a discussion on likelihood definition and parameter space exploration procedure in Section \ref{sec:params_explore}. Next, in Section \ref{sec:results} we discuss our results of forecast using different FRB samples followed by the inclusion of 21~cm power spectra information as well.  Finally, we summarize our main findings in Section \ref{sec:conc}. In this paper, the assumed cosmological parameters are $\Omega_M$ = 0.308, $\Omega_{\Lambda}$ = 0.691 $\Omega_b$ = 0.0482, $h$ = 0.678, $\sigma_8$ = 0.829 and $n_s$ = 0.961 \citep{2016A&A...594A..13P}.

\section{Reionization Model}
\label{sec:theory}

 The reionization models exploited in \texttt{SCRIPT} are discussed in earlier studies \citep{ 2022MNRAS.511.2239M, 2022MNRAS.515..617M} with details. In this section, we provide a brief overview for the completeness of the discussion which closely follows \citet{2023MNRAS.526.3920M}.

\texttt{SCRIPT} provides the ionization state of the universe in a cosmologically representative simulation volume which further enables us to utilize converged large scale power spectra of ionization fluctuation with respect to the resolution of simulation box \citep{2018MNRAS.481.3821C}. To initiate the model, we provide inputs comprising the density field and the distribution of collapsed halos capable of emitting ionizing radiation. Given our focus on large-scale features of the IGM, we employ the second-order Lagrangian Perturbation Theory (2LPT) approximation \citep{1998MNRAS.299.1097S} to generate the density field, rather than conducting a full N-body simulation. Specifically, we utilize the implementation by \citet{2011MNRAS.415.2101H}.\footnote{\url{https://www-n.oca.eu/ohahn/MUSIC/}}. To compute the halo distribution, we rely on a sub-grid prescription, based on the conditional ellipsoidal mass function \citep{2002MNRAS.329...61S}. In this study, we utilize a simulation box size of $256h^{-1}\mathrm{cMpc}$, which proves to be sufficient for the observables relevant to our analysis as recent investigations have demonstrated \citep{2014MNRAS.439..725I,2020MNRAS.495.2354K}.
 We use a resolution of $\Delta x=8~h^{-1}\mathrm{cMpc}$ allowing for efficient model computation while capturing the inhomogeneous effects during reionization. For our model to compute the complete reionization history, we utilize comoving simulation boxes spanning from $z = 20$ to $z = 5$, with an interval of $\Delta z=0.1$.

Our model utilizes the photon-conserving algorithm to construct the reionization topology within the simulation box. The ionization field relies on the ionization efficiency parameter $\zeta(M_h, z)$, which estimates the available ionizing photons per hydrogen atom. This parameter may vary based on halo mass ($M_h$) and redshift ($z$). More specifically, this parameter provides an estimate of the available number density of ionizing photons ($n_{\mathrm{ion},i}$ in each cell $i$ of the simulation box which can be compared against the available number density of hydrogen atoms in that cell, $n_{H,i}$. To address inhomogeneous recombinations, we adjust the ionization criteria compensating for excess neutral atoms ($n_{\mathrm{rec},i}$) i.e. at each step of the algorithm, we flag a cell as ionized if $n_{\mathrm{ion},i} \ge n_{H,i} + n_{\mathrm{rec},i}$ and distribute the excess photons to the nearby cells. The cells failing to satisfy the criterion are assigned with an ionization fraction of $(n_{\mathrm{ion},i}-n_{\mathrm{rec},i})/n_{H,i}$. To compute the recombination number density we need to track the density and ionization evolution self consistently, while introducing small-scale fluctuations through a globally averaged clumping factor, $C_{\mathrm{HII}}$. This parameter remains highly uncertain due to the absence of direct observational measurements, although previous simulation studies have hinted at a value around $2-5$ \citep[e.g.][]{2020ApJ...898..149D}. In this study, we keep this parameter fixed throughout with a value of  $3$. The computation of recombination number density also requires information of the thermal evolution. Hence, in addition to tracing the ionization history, we compute the thermal evolution of each grid cell within the box. The code automatically incorporates the impact of spatially varying reionization on temperature evolution. It assumes that a region's temperature increases by a value, $T_{\mathrm{re}}$, as it undergoes ionization for the first time. This parameter, $T_{\mathrm{re}}$, is commonly referred to as the reionization temperature \citep{1997MNRAS.292...27H,2009ApJ...701...94F,2018MNRAS.477.5501K,2022MNRAS.511.2239M}.

Our methodology also incorporates radiative feedback, which suppresses the generation of ionizing photons within halos where the gas is heated. In \citet{2022MNRAS.511.2239M} several approaches were introduced to integrate radiative feedback effects. Here, we adopt the `step feedback' model, where the gas fraction retained within halos affected by radiative feedback is assumed to be zero for halo masses smaller than $M_{\mathrm{min}} = \mathrm{Max} \left[M_{\mathrm{cool}}, M_{J}\right]$, and unity otherwise. Here, $M_{\mathrm{cool}}$ represents the minimum threshold mass for atomic cooling, while $M_J$ denotes the Jeans mass at virial overdensity. This method provides an efficient and straightforward way of incorporating radiative feedback during reionization. $M_{\mathrm{min}}$ varies not only with redshift but also spatially, particularly when the feedback effect dominates within ionized regions. This spatial variation arises because $M_J$ is temperature-dependent, and different regions exhibit distinct temperatures following the ionization topology. However, in neutral regions unaffected by feedback, $M_{\mathrm{min}}$ is determined by the threshold for atomic cooling mass (i.e. $M_{\mathrm{cool}}$), which is approximately $\sim 3\times 10^7 M_{\odot}$ at $z = 20$ and $\sim 2\times 10^8 M_{\odot}$ at $z = 5$. In regions affected by feedback, $M_{\mathrm{min}}$ typically exceeds $10^9 M_{\odot}$. While more sophisticated and realistic models exist, where the feedback effect is gradual and assumes a mass-dependent depletion of gas fraction instead of a step-like cutoff, these implementations are computationally less efficient compared to the `step feedback' model. Therefore, for the purpose of this study, employing a step-like feedback approach can be considered reasonable.

We list the free parameters of our model below: 

\begin{itemize}
\item We assume an ionization efficiency parameter $\zeta$, which is independent of halo mass $M_h$ but varies with redshift. This choice is motivated by the earlier works \citep{2015ApJ...813...54T,2016MNRAS.460..417S,2020MNRAS.495.3065D,2021MNRAS.501L...7C,2022MNRAS.511.2239M}, where the redshift-dependence is assumed to be a simple power-law
\be
\label{eq:zeta_eq}
    \zeta(z) = \zeta_0\left(\frac{10}{1+z}\right)^\alpha,
\ee
with $\zeta_0$ being the ionization efficiency at $z = 9$ and $\alpha$ is the slope. We use $\log \zeta_0$ and $\alpha$ as free parameters in our models. 

\item As discussed earlier, the reionization temperature $T_{\mathrm{re}}$ is crucial for the IGM temperature evolution, recombination, and radiative feedback effects. We use $\log T_{\mathrm{re}}$ as a free parameter while parameter exploration.
\end{itemize}

So, we have three free parameters $\mathbf{a} = \{\log \zeta_0, \alpha, \log T_{\mathrm{re}}\}$ for the study presented in this paper. Along with these, we also derive constraints on some of the parameters which can be crucial to characterize the timeline for epoch of reionization. These derived parameters are CMB scattering optical depth ($\tau_e$), the reionization interval between 25\% to 75\% ionization ($\Delta z$) and the midpoint of reionization ($z_{\mathrm{mid}}$), following \citet{2022MNRAS.515..617M}. 

\begin{figure*}[tbp]
\centering 

\subfigure[DM-$z$ variation for different $\zeta_0$ models ]{ \includegraphics[width=\columnwidth]{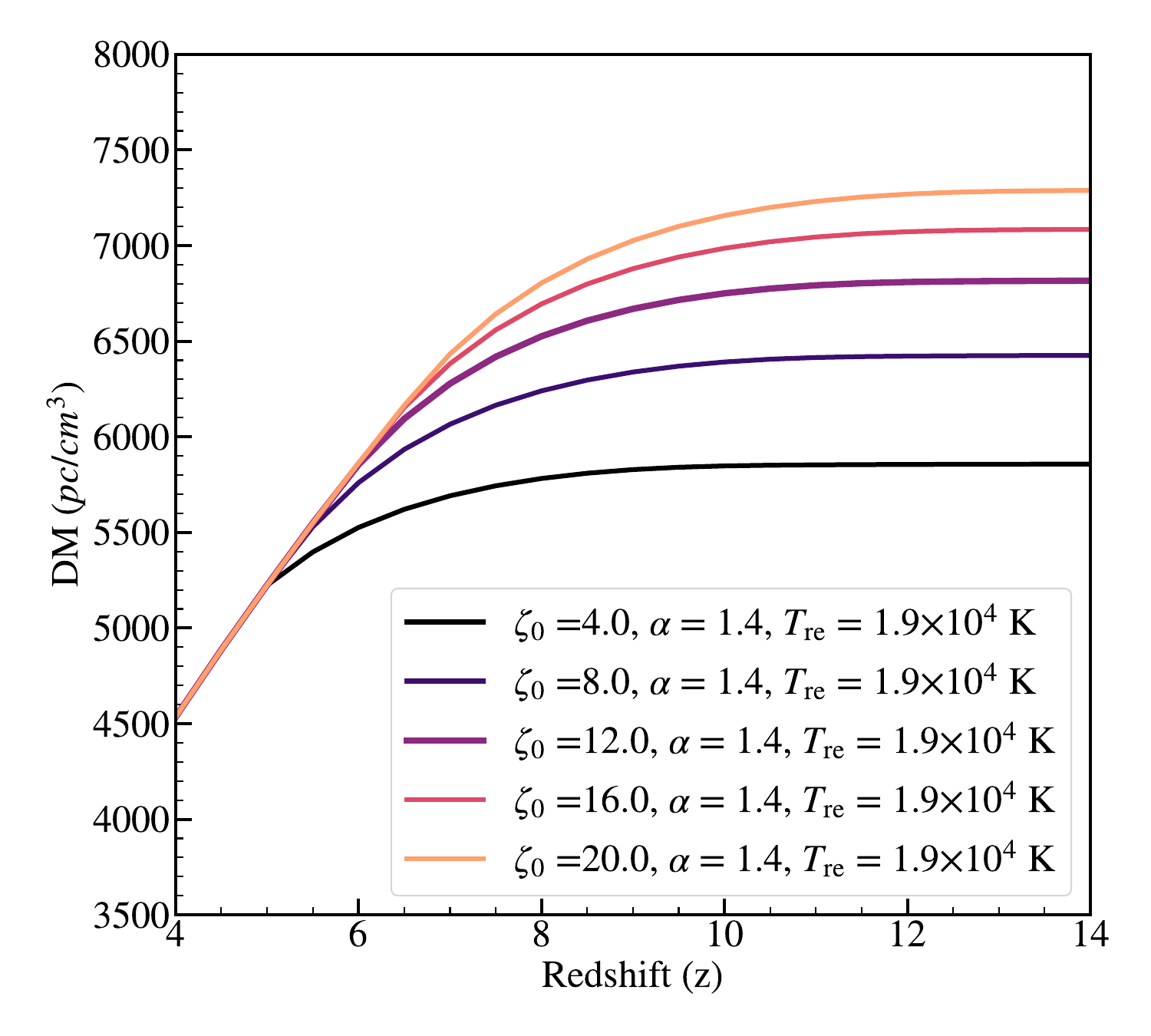}}
\subfigure[DM-$z$ variation for different $\alpha$ models ]{\includegraphics[width=\columnwidth]{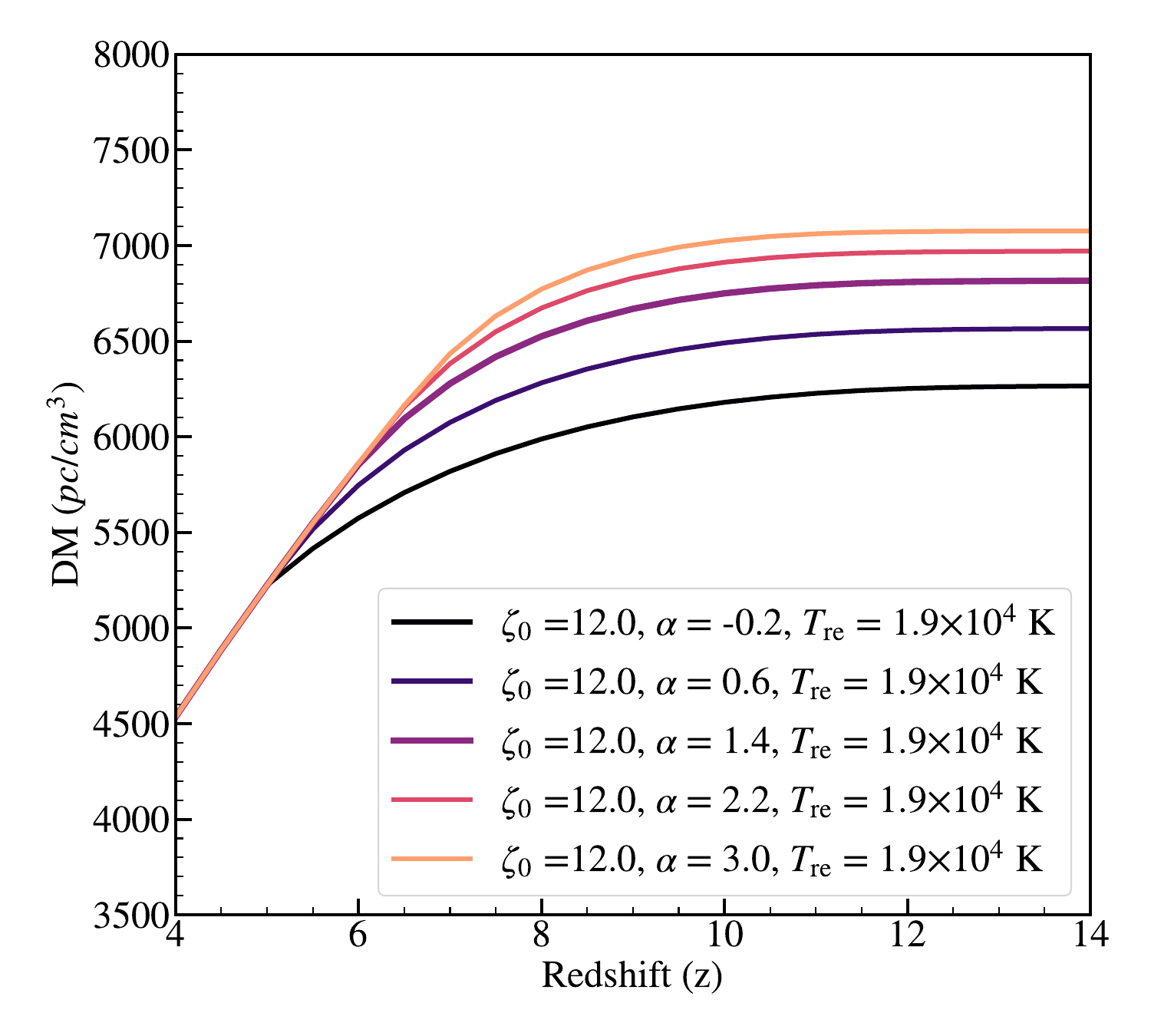}}
\subfigure[DM-$z$ variation for different $T_{\mathrm{re}}$ models ]{\includegraphics[width=\columnwidth]{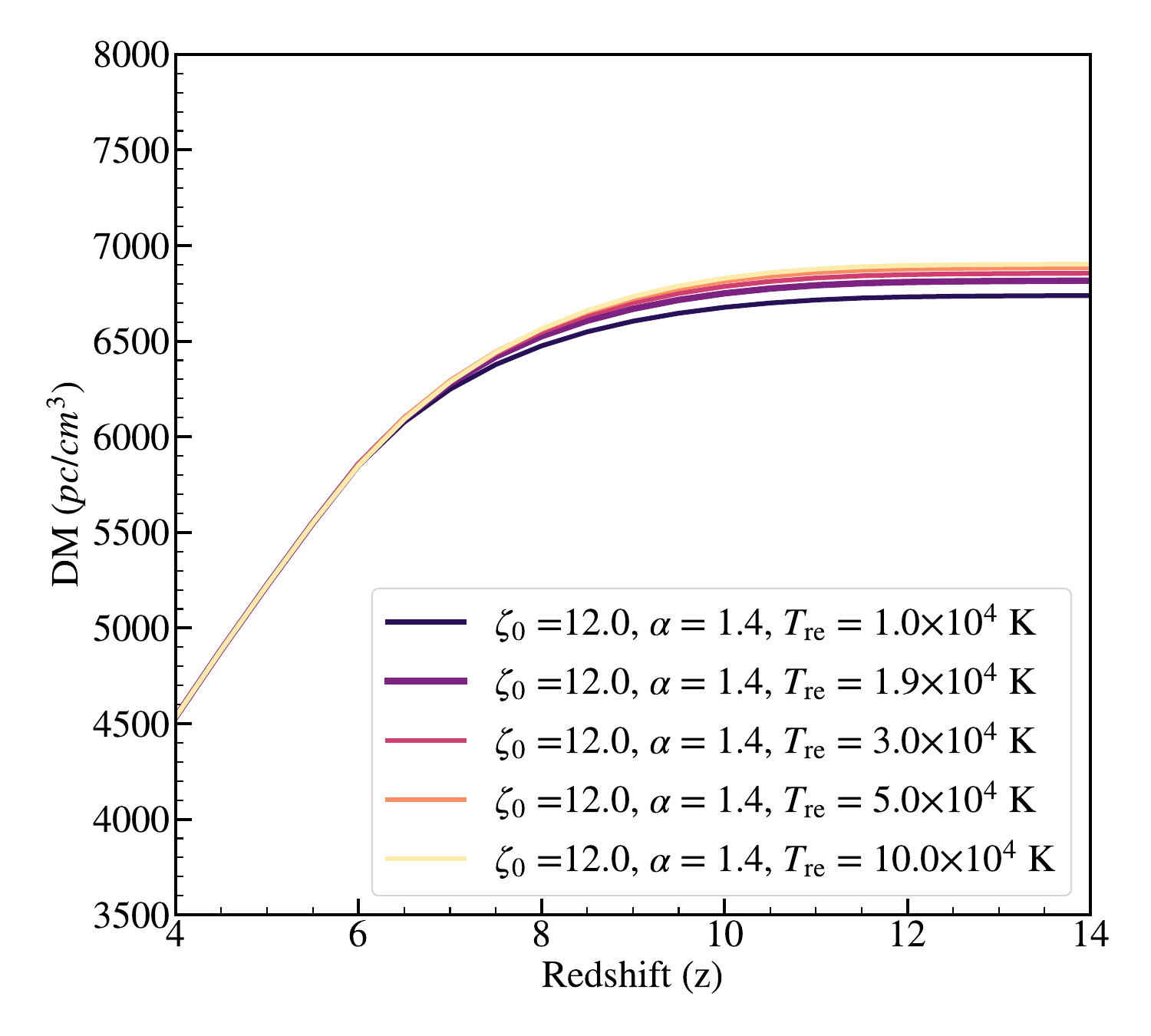}}
\subfigure[Mock data with 100 samples using fiducial model]{\includegraphics[width=\columnwidth]{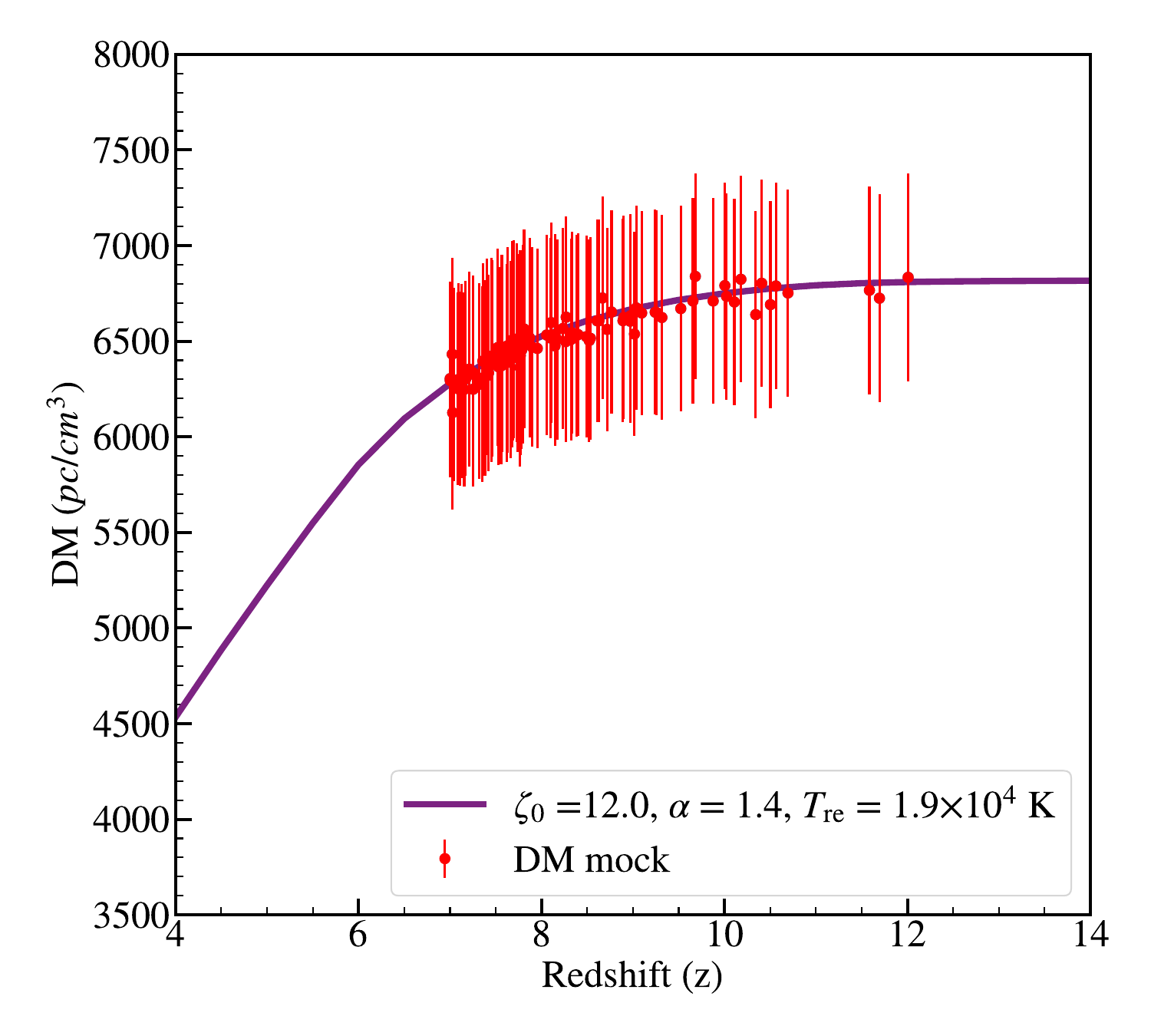}}
\caption{\label{fig:DM_with_param} The first three panels (a,b,c) show the variation of globally averaged Dispersion Measure (DM) as a function of redshift for different free parameters in our model (i.e. $\zeta_0$, $\alpha$ and $T_{\mathrm{re}}$). We use $\zeta_0=12$, $\alpha=1.4$, and $T_{\mathrm{re}}=1.9\times10^4~ \mathrm{K}$ as our fiducial model parameters in the analysis later. For each case, we show the plots for five different parameter values. The last panel (d) shows the fiducial mock DM values (at true redshifts) for 100 samples along with the corresponding input model. The mock data have been generated assuming a typical expectation from upcoming facilities like SKA.}
\end{figure*}


\section{Dispersion Measure (DM) of FRBs}
\label{sec:DM_FRB}
Dispersion Measure (DM) is an estimate of the integrated column density of the free electrons residing in the medium between an observed astronomical source and us. For high redshift FRBs, DM is constituted of mainly three parts through which the photons originated from the bursts traverse towards us. These include the contribution from the FRB host galaxy, the IGM, and Milky Way i.e.
\be
    \mathrm{DM = DM^{\mathrm{host}} + DM^{\mathrm{IGM}} + DM^{\mathrm{MW}}}
\ee

The corresponding uncertainties can be added in quadrature to get an overall estimate of DM uncertainties.
\be
\label{eq:sigma_tot}
\sigma_{\mathrm{DM}} = \sqrt{\left(\sigma_{\mathrm{DM}}^{\mathrm{host}}\right)^2 + \left(\sigma_{\mathrm{DM}}^{\mathrm{IGM}}\right)^2 + \left(\sigma_{\mathrm{DM}}^{\mathrm{MW}}\right)^2}
\ee
Another source of uncertainties comes from the observation itself. As we will discuss later in section \ref{sec:generate_mock}, we take that into account while generating the mock dataset. Next, we discuss the procedure for estimating DM accounting for the various contributions mentioned above.

\subsection{IGM contribution} The major contribution for high redshift FRBs comes from the free electrons present in the IGM. For a FRB at redshift $z$, the mean IGM contribution to DM is given by 
\be
\langle \mathrm{DM}\rangle^{\mathrm{IGM}} = \int_0^z \f{c\langle n_e(z')\rangle}{H(z')(1+z')^2}\de z'
\ee
where $c$ is the speed of light in vacuum, $H(z')$ is the Hubble parameter at redshift $z'$ and $n_e$ is the mean IGM electron density. $\langle...\rangle$ denotes average over the whole volume of the simulation box. The above expression can further be written in terms of astrophysical and cosmological parameters as
\be
\label{eq:DM_eq}
\langle \mathrm{DM}\rangle^{\mathrm{IGM}} = \f{c}{H_0}\f{(1-Y)f_{\mathrm{IGM}}\Omega_bh^2\rho_{c,0}}{m_{\mathrm{H}}}\int_0^z\f{\chi_{\mathrm{He}}(z')Q_{\mathrm{HII}}^M(z')(1+z')}{E(z')}\de z'
\ee
where $m_{\mathrm{H}}$ is the proton mass, $Y$ is the helium mass fraction, $\rho_{c,0}$ is the critical density, and $\chi_{\mathrm{He}}$ takes into account for the excess electrons arising from helium reionization. $Q_{\mathrm{HII}}^M = \langle x_{\mathrm{HII},i}\Delta_i\rangle$ is the mass averaged ionization fraction where $x_{\mathrm{HII},i}$ and $\Delta_i$ are the ionization fraction and the overdensity of any $i$-th cell in our simulation box respectively.
The quantity $f_{\mathrm{IGM}}$ takes into account the fact that FRBs trace the electron distribution in the IGM density instead of cosmic mean density which is slightly larger due to the electron contribution from high density environments. However, this value is actually close to unity in the case of high redshift  FRBs. In this study, we take it to be 1 as we will be dealing with redshift $z>6$ physics \citep{2022ApJ...933...57H}.

The uncertainty in the DM contribution from IGM is estimated from an empirical relation based on low redshift FRB analysis \citep{2019PhRvD.100h3533K}
\be
\sigma_{\mathrm{DM}}^{\mathrm{IGM}}(z) = \frac{0.2}{\sqrt{z}} \langle \mathrm{DM}\rangle^{\mathrm{IGM}}
\ee
We use an extrapolation of this relationship for higher redshifts. As implemented in earlier studies, the extrapolated maximum uncertainty is at $z\sim6.5$. Hence, we assume $\sigma_{\mathrm{DM}}^{\mathrm{IGM}}(z>6.5)=\sigma_{\mathrm{DM}}^{\mathrm{IGM}}(6.5)$ for the higher redshifts, following the literature \citep{2021MNRAS.502.2346H}. As the relation is empirical, it can be modified a bit with the availability of more FRBs. However, those changes are not expected to alter the statistical analysis significantly.
\subsection{Galactic (host + Milky Way) contribution} As the DM is a line of sight integrated quantity, it increases with the distance of the object. In case of high redshift FRBs (hence at a longer distance), the DM contribution from our Milky Way galaxy and FRB host galaxy are expected to be much smaller than the IGM.  Although the host contribution is not well determined, several studies have proposed this to be $\mathrm{DM}^{\mathrm{host}} \lesssim 200~ \mathrm{pc/cm^3}$. For Milky Way contribution, the CHIME/FRB catalogue has estimates to be $\mathrm{DM}^{\mathrm{MW}} \lesssim 500~ \mathrm{pc/cm^3}$ \citep{2021ApJ...922...42R}. Following \citet{2022ApJ...933...57H}, we adopt a conservative value of $\sim 750~\mathrm{pc/cm^3}$ accounting for any other less significant contribution to the DM.

As discussed earlier, these DM estimates are associated with large uncertainties which is also true for host contribution as well \citep{2017ApJ...834L...7T}. Hence, we adopt a relatively large value for host DM uncertainties of $\sigma_{\mathrm{DM}}^{\mathrm{host}} = 100 ~\mathrm{pc/cm^3}$ following \citet{2021JCAP...05..050D}. Similarly, we assume an uncertainty of  $\sigma_{\mathrm{DM}}^{\mathrm{MW}}\sim 54~\mathrm{pc/cm^3}$ \citep{2022ApJ...933...57H} which takes the modelling uncertainty as well as measurement uncertainty into account.

\section{DM variations with parameters}
\label{sec:DM_variation}
In this section, we discuss how DM distribution with redshift behaves for different parameters in our models.

From equation \ref{eq:DM_eq}, it is straightforward to say that DM increases as redshift becomes higher. DM distribution is expected to depend upon the reionization history of the universe. An early reionization end can produce a significant amount of free electrons in the IGM which shows up a relatively higher DM. Similarly, for a late reionization end, the DM values are expected to be lower.

In Figure \ref{fig:DM_with_param}, we show the redshift evolution of global DM values. At the \textit{top left} (panel a), we find that a higher value of $\zeta_0$ produces a higher DM distribution. This is expected as high $\zeta_0$ provides an early reionization end due to efficient photon production which eventually contributes to boosting up the DM values. For a similar reason, at the \textit{top right} panel (panel b), the enhancement in $\alpha$ values also provides high global DM. At the \textit{bottom left} (panel c), we show the DM variation with the temperature increment $T_{\mathrm{re}}$ parameter. In this case, the variation is comparatively weaker than in the previous two cases. The higher $T_{\mathrm{re}}$ values generate slightly higher DM distributions. This is because a high temperature model reduces the recombination rate (as the recombination coefficient decreases with temperature) and hence, reionization proceeds faster at early stages. However, at later stages, this effect is compensated when radiative feedback starts to dominate \citep{2022MNRAS.511.2239M}. The balance between these two opposite trends sets the reionization evolution and, hence the DM distribution.

From this analysis, it is clear that the FRBs at high redshifts ($z>7$) can play a crucial role in constraining ionization and thermal evolution of the IGM. Similar to the 21~cm signal, at present, there is not yet any detection of FRBs at these redshifts. So, we pursue the forecasting study with realistic mock which are expected to be available in the near future when the major observational facilities will come online.
\begin{figure*}
    \centering
    \includegraphics[width=0.8\textwidth]{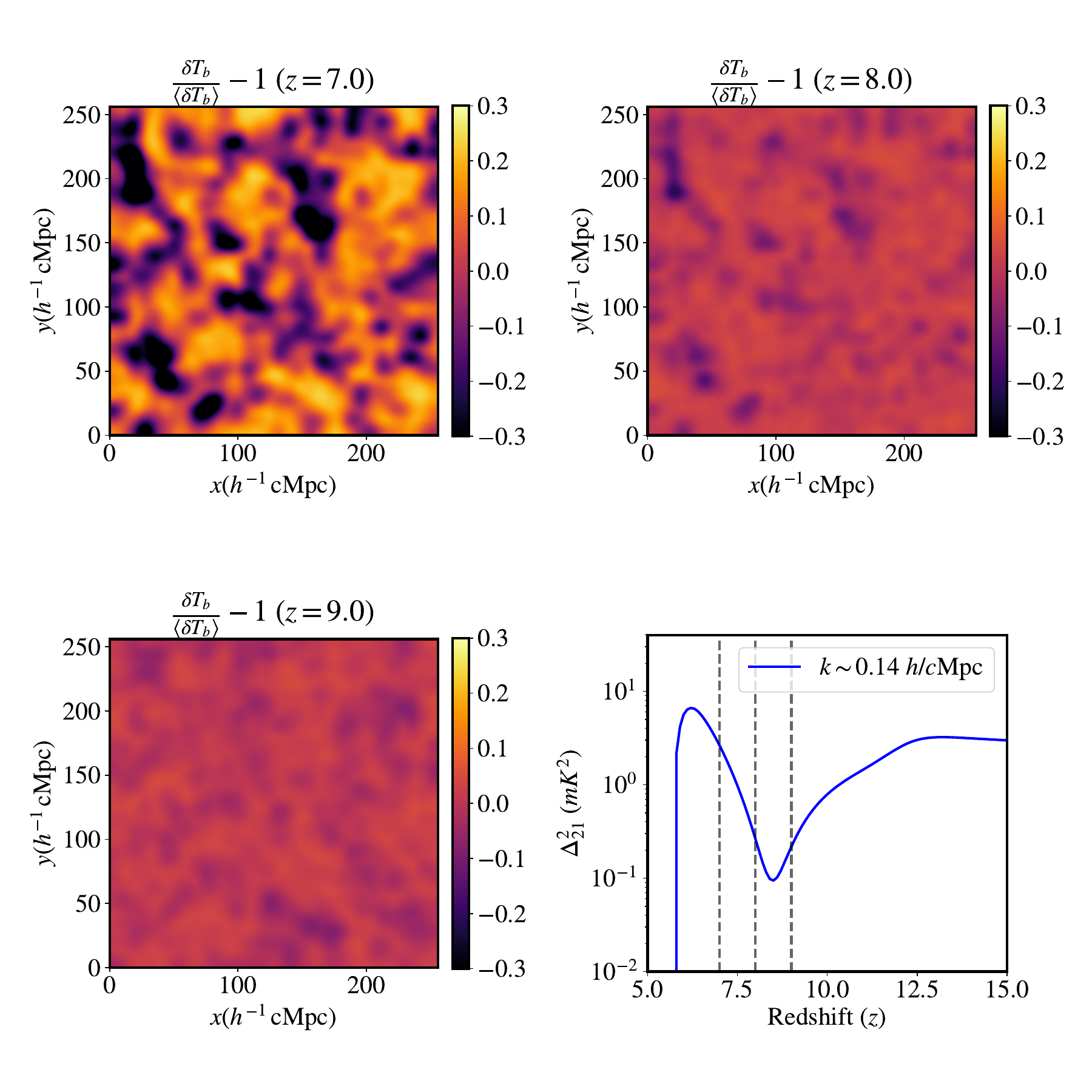}
    \caption{The snapshots of normalized mean subtracted fluctuation in 21~cm field at three different redshifts for our fiducial set of model parameters which has been used to generate the mocks ($z=7.0,~ 8.0$ at the \textit{top} and $z=9.0$ at the \textit{bottom left}) along with the redshift evolution of large scale ($k\sim0.14~h/c\mathrm{Mpc}$) 21~cm power spectra (at the \textit{bottom right}). The dashed lines correspond to the redshifts which we utilize in the likelihood analysis.}
    \label{fig:fid_21}
\end{figure*}
\section{Generating Mock Data}
In this section, we discuss the procedure of generating a realistic mock dataset for the forecasting study. We generate the mock assuming a fiducial reionization model given by the parameters  $\zeta_0=12~(\log \zeta_0\sim 1.079)$, $\alpha=1.4$, and $T_{\mathrm{re}}=1.9\times10^4~ \mathrm{K}~ (\log T_{\mathrm{re}}\sim 4.279)$. This provides a realistic reionization history obeying a variety of observables (including UV luminosity function, low density IGM temperature estimates, and CMB scattering optical depth) as motivated from an earlier parameter exploration study \citep{2023MNRAS.526.3920M}.
\label{sec:generate_mock} 
\subsection{FRB DM mocks}
In this work, we do the forecast with three sets of samples containing $100$, $500$, and $1000$ FRBs.
As we are interested in the EoR, we generate these samples for a redshift range of $z=7-15$. Here, we adopt the fact that the number of FRBs follows the star formation rate which has been observed at low redshifts \citep{2021MNRAS.501.5319A,2022MNRAS.510L..18J}. Following the method described in \citet{2022ApJ...933...57H}, we assume the empirical scaling relation for cosmic star formation rate (SFR) provided by \citep{2019MNRAS.488.3143B}.  This provides around 80\% of samples within $z<10$. We adopt this conservative approach as the prediction of star formation rate at high redshifts is highly uncertain. With the advent of sensitive telescopes like JWST, TMT, it is possible to get a better estimate of high redshift SFR which can be different than the adopted fit. In that case, the derived constraints on reionization models are expected to be stronger with higher SFR and weaker with lower SFR as the number of high redshift samples depends on the SFR.  

Once the redshift samples are generated, we add the observational noise assuming uncertainties of 10\% on the spectroscopic data expected from telescopes like JWST. So, to get the observational redshifts $z_{\mathrm{obs}}$, we draw the numbers from a Gaussian random distribution with mean at true redshift value ($\mu =z_{\mathrm{tr}}$) and standard deviation of $\sigma_z= 0.1z_{\mathrm{tr}}$. Next, we compute the total DM contributions and uncertainties ($\sigma_{\mathrm{DM}}$) for each sample according to the equations described in section \ref{sec:DM_FRB}. We evaluate the IGM contributions for the true redshift values and add the host halo and Milky Way contributions on top of that. To account for the measurement uncertainties, we shift these DM values with a Gaussian random number of mean zero and width of $50$, which can then be treated as observational mock ($\mathrm{DM^\mathrm{obs}}$). This width is motivated by the analysis presented in \citet{2022ApJ...933...57H}, where it has been shown that measurement uncertainties remain $<100~ \mathrm{pc/cm^3}$ for an instrument like FAST and it is expected that upcoming telescope like SKA will provide more stringent measurement. The DM uncertainties due to host, IGM, and Milky Way are estimated as discussed earlier and the total observational uncertainties are computed using equation  \ref{eq:sigma_tot}.

In the last panel (\textit{bottom right}, d) of figure \ref{fig:DM_with_param}, we show the mock DM values generated with 100 FRB samples along with corresponding uncertainties. We also show the actual DM evolution for the underlying input reionization model. It is apparent that the density of data points decreases as we move towards the higher redshifts which is a direct consequence of our assumption on redshift evolution of cosmic the star formation rate.
\subsection{21 cm mocks}
As our semi numerical reionization model provides the ionization field at the redshift of interests, we can compute the 21 cm differential brightness temperature for each cell in the simulation box which is given by \citep{1997ApJ...475..429M,2003ApJ...596....1C}
\be
\label{eq:delta_Tb}
\delta T_{b, i} \approx 27~\mathrm{mK}  \left(1 - x_{\mathrm{HII}, i}\right) \Delta_i \left(\frac{1+z}{10}\frac{0.15}{\Omega_{m}h^2}\right)^{1/2} \left(\frac{\Omega_{b}h^2}{0.023}\right),
\ee
where $\Delta_i \equiv \rho_{m, i} / \bar{\rho}_m$ is the ratio of the matter density $\rho_{m,i}$ in the grid cell and the mean matter density $\bar{\rho}_m$.

The observable we explore here is the dimensionless 21~cm power spectrum, defined as
\begin{equation}
\label{eq:Delta_21}
   \Delta_{21}^2(k) = \frac{k^3 P_{21}(k)}{2 \pi^2},
\end{equation}
where $P_{21}(k)$ is the power spectrum of the mean-subtracted fluctuation field $\delta T_{b,i} - \langle \delta T_{b, i} \rangle$.

As we are dealing with a relatively coarse resolution box, we only consider the large scale modes of the power spectra, specifically $k\sim0.14~h/\mathrm{cMpc}$ where the upcoming interferometric data from telescopes like SKA-low, HERA are expected to provide the detection (even better than a percentage level with sufficient observational time). 
This choice of a single $k$ mode has been kept a bit conservative as including more high $k$ modes can be affected by the lack of resolution while the lower $k$ modes are generally dominated by the cosmic variance uncertainties. For this analysis, we choose three suitable redshifts (i.e. $z=7.0,8.0~\&~9.0$) covering different phases of the ionization state. To generate the 21 cm mocks, we use the theoretical power spectrum for our fiducial model parameters (as discussed earlier) and add a random number generated from a Gaussian distribution with zero mean and standard deviation equal to the thermal noise of the specific bin. We calculate the thermal noise using a modified version \texttt{21cmsense} code \citep{2013AJ....145...65P,2014ApJ...782...66P} assuming SKA-low specifications \citep{2019arXiv191212699B}. We further assume the noise for observations of $\sim 1080$ hours along with moderate foreground removal \citep{2014ApJ...782...66P}. For the total noise,  we take a conservative/pessimistic approach and assume an expected uncertainty which is 10\% of the theoretical power spectra. This is slightly larger than the expected uncertainty after adding cosmic variance contribution, however, this is reasonable enough to serve the purpose as a proof of concept study. To correctly account for the cosmic variance contribution, we generate the mock data with a random initial seed (for density distribution) different from the one utilized in the parameter space exploration study discussed later. 

In Figure \ref{fig:fid_21}, we show the normalized 21~cm fluctuation maps using our fiducial model at the earlier mentioned redshifts. It can be seen that the fluctuations increase as we move from the initial stages ($z\sim8.0, 9.0$, \textit{top right, bottom left}) to the middle stage of reionization ($z\sim7.0$, \textit{top left}). This is not surprising as the percolation between the ionized regions start to begin around the middle stages of reionization. A similar trend can be followed from the amplitudes of the power spectra (\textit{bottom right}) at a $k$ value of $0.14~h/c\mathrm{Mpc}$ which basically quantify the amount of these fluctuations at a large scale. The dashed vertical lines correspond to the three redshifts utilized in the likelihood analysis. At a very initial stage, the power spectrum is dominated by the density field fluctuations (note that our models don't include any heating during cosmic dawn as our focus is EoR in this study). As the high density regions ionize first, this neutralizes the 21 cm fluctuations a bit, hence, suppressing the amplitude of the power spectra. Then, it starts increasing when ionization starts to dominate (around $z\sim 8-9$).
\section{Parameter Space Exploration}
\label{sec:params_explore}

We utilize standard Bayesian approach to explore the parameter spaces. Our aim is to compute the conditional probability distribution or the posterior $\mathcal{P}(\lambda \vert \mathcal{D})$ of the model parameters $\lambda$, provided the observational (i.e. mock in this study) data sets $\mathcal{D}$ mentioned in the previous section. This can be computed using the Bayes theorem
\be\label{eq:bayes_eq}
 \mathcal{P}(\lambda\vert \mathcal{D})=\frac{\mathcal{L}(\mathcal{D} \vert \lambda) ~\pi(\lambda)}{\mathcal{P}(D)},
\ee
where $\mathcal{L}(\mathcal{D} \vert \lambda)$ is the conditional probability distribution of data given the parameters or the likelihood, $\pi(\lambda)$ is the prior and $\mathcal{P}(\mathcal{D})$ is the evidence (which can be treated as the normalization parameter and does not play any role in our analysis). In our case, $\mathcal{D}$ is the dataset consisting of $\mathrm{DM}^{\mathrm{obs}}$ and $z^{\mathrm{obs}}$. The likelihood is chosen to be multi-dimensional Gaussian
\bear
\label{eq:chisq_eq}
\mathcal{L_{DM}}( \lambda) 
&= \exp \left(-\frac{1}{2} \sum_{j}\left[\frac{\mathrm{DM}^{\mathrm{obs}}_j-\mathrm{DM}^{\mathrm{model}}_j(\lambda)}{\sigma_{\mathrm{DM},j}}\right]^2 \right)
\nline
&\times  \exp \left(-\frac{1}{2} \sum_{j}\left[\frac{z^{\mathrm{obs}}_j-z^{\mathrm{tr}}_j}{\sigma_{z,j}}\right]^2 \right),
\ear
where $\mathrm{DM}^{\mathrm{obs}}_j$ are the mock data points, $\mathrm{DM}^{\mathrm{model}}_j(\lambda)$ are the predictions for DM values using the parameter set $\lambda$ and $\sigma_{\mathrm{DM},j}$ are the estimated observational uncertainties on the mock data as discussed previously in Section \ref{sec:generate_mock}. The summation index $j$ runs over all data points used in the analysis. For 21 cm mock data, the likelihood is defined as 
\bear
\label{eq:chisq_eq_21}
\mathcal{L}_{21}(\lambda) 
&= \exp \left(-\frac{1}{2} \sum_{\alpha,\beta}\left[\frac{\Delta_{21,\mathcal{D}}^2(k_{\alpha},z_{\beta})-\Delta_{21}^2(k_{\alpha}, z_{\beta}; \lambda)}{\delta\Delta_T^2(k_{\alpha},z_{\beta})}\right]^2 \right)
\ear
where $\Delta_{21}^2(k_{\alpha},z_{\beta}; \lambda)$ are the model predictions for the parameters $\lambda$ at a redshift ($z_{\beta}$) and $k$ bin ($k_{\alpha}$), $\Delta_{21,\mathcal{D}}^2(k_{\alpha},z_{\beta})$ are the mock 21~cm data points and $\delta \Delta_T^2(k_{\alpha},z_{\beta})$ are the corresponding error bars on the data. The summation index $\beta$ runs over all redshift values used in the analysis and $\alpha$ is just a single value in our case. The joint likelihood using FRB and 21 cm data can be derived as $\mathcal{L}(\lambda) = \mathcal{L_{DM}}( \lambda)\times \mathcal{L}_{21}(\lambda)$. 

Along with this, we include two observationally motivated priors on the ionization evolution history of the universe. We assume that the reionization process is end ($Q_{\mathrm{HI}}^V\le 0.01$) by redshift $z=5.3$ which has been suggested by recent works based on Ly-$\alpha$ forest observations available from high redshift quasar spectra \citep{2022MNRAS.514...55B}.  Also, we assume a conservative prior that the universe can not be fully neutral at a redshift $z\le10$, motivated by the limits put on neutral fraction with recent JWST discovery of high redshift Lyman-break galaxy \citep{2023ApJ...949L..40B}. We also do not include exotic reionization models showing sudden ionization or doubly peaked ionization.

We sample the posterior distribution using the Monte Carlo Markov Chain (MCMC) method, more specifically, the Metropolis-Hastings algorithm \citep{1953JChPh..21.1087M}. We utilize the publicly available package \texttt{cobaya} \citep{2021JCAP...05..057T}\footnote{\url{https://cobaya.readthedocs.io/en/latest/}} to run the MCMC chains.  
We check the convergence of the chains following the  Gelman-Rubin $R - 1$ statistic \citep{1992StaSc...7..457G}. The chain is assumed to be converged if the $R - 1$ value becomes less than a threshold $0.01$. 
For inference, we remove the first $30\%$ of samples from the chains as `burn-in' and study with the rest which is critical to avoid spurious correlations among the parameters. 
\section{Results}
\label{sec:results}
\begin{figure*}
    \centering
    \includegraphics[width=0.7\textwidth]{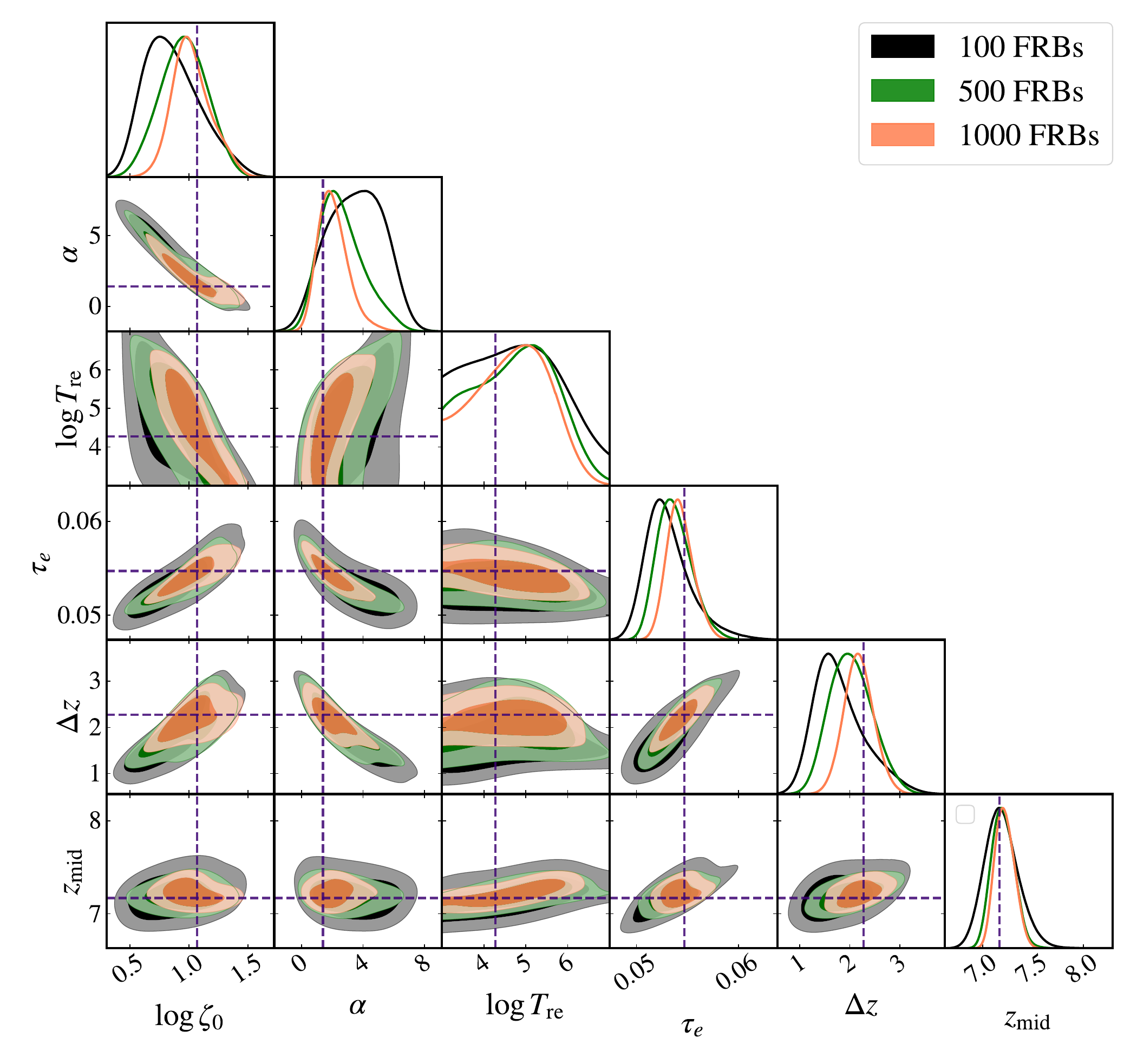}
    \caption{The comparison of parameter posteriors for three different sets of FRB mocks, i.e. with 100 (\textit{black}), 500 (\textit{green}) \& 1000 (\textit{orange}) samples. The diagonal panels show the 1D posterior probability distribution while off diagonal panels show the joint 2D posteriors. The contours represent 68\% and 95\% confidence intervals. The dashed line represents the input parameter values used to generate the reionization model which has further been used to create the FRB mocks. }
    \label{fig:corner_constr}
\end{figure*}
\begin{figure*}
    \centering
    \includegraphics[width=0.85\textwidth]{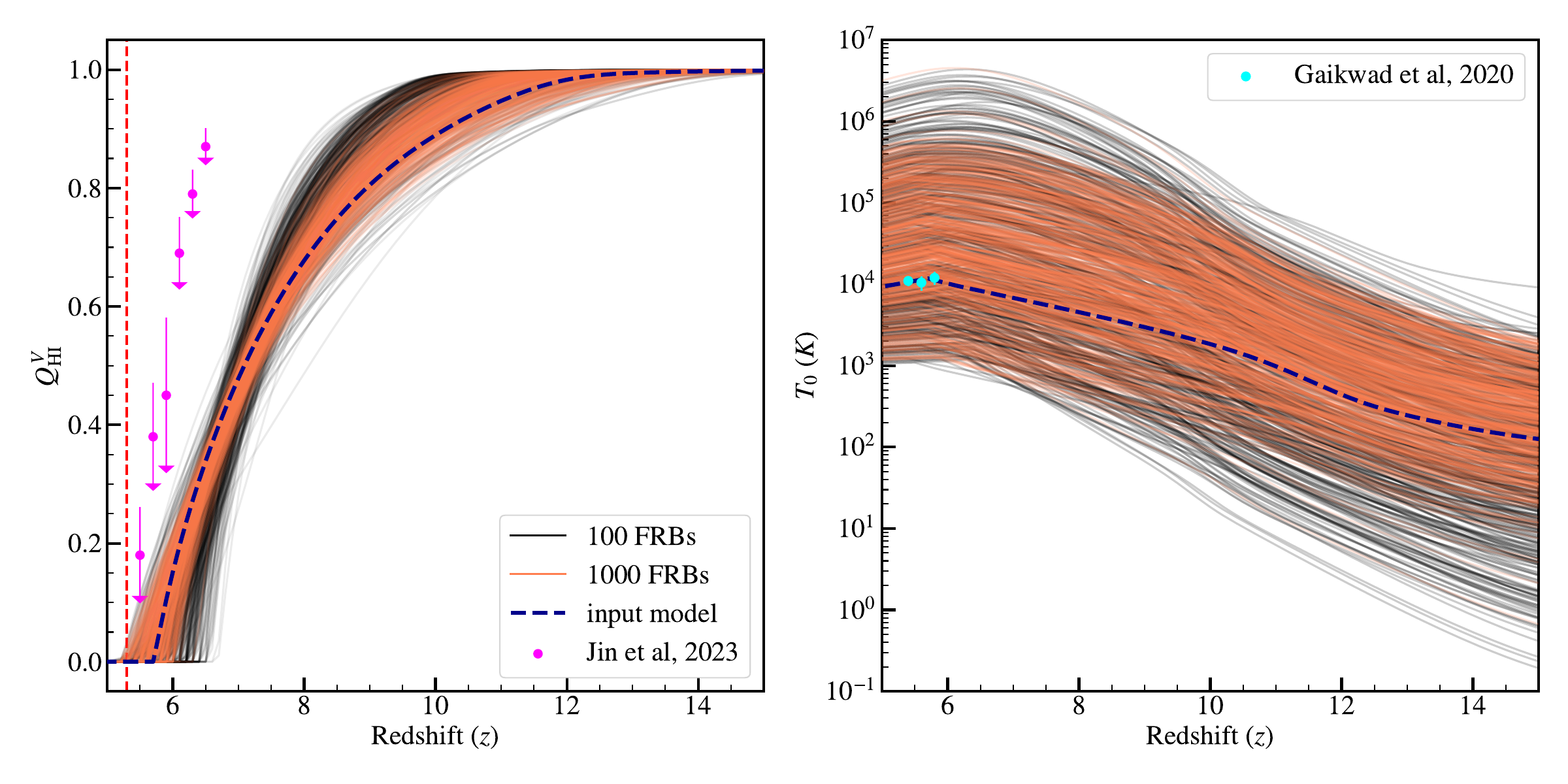}
    \caption{Plots of 500 random models from the posterior chains for two different sets of FRB mocks (with 100 : \textit{black}; with 1000 : \textit{orange}). At the \textit{left} panel: we show the evolution of neutral hydrogen fraction with redshift. The red vertical dashed line represents our hard prior on the end of reionization at $z=5.3$. At the \textit{right} panel: we show the mean IGM temperature evolution. Although we don't use any other observational constraints while pursuing parameter space exploration, we show some of the existing observational estimates for ease of comparison. 
    The magenta points at the \textit{left} represent the recent upper limits on neutral fraction from \citep{2023ApJ...942...59J}. The cyan points at the \textit{right} are the constraints on the mean IGM temperature using Ly-$\alpha$ forest spike statistics study \citep{2020MNRAS.494.5091G}. The blue dashed line represents the input reionization model used for generating the FRB mocks. }
    \label{fig:posterior_models}
\end{figure*}
\begin{figure*}
    \centering
    \includegraphics[width=0.7\textwidth]{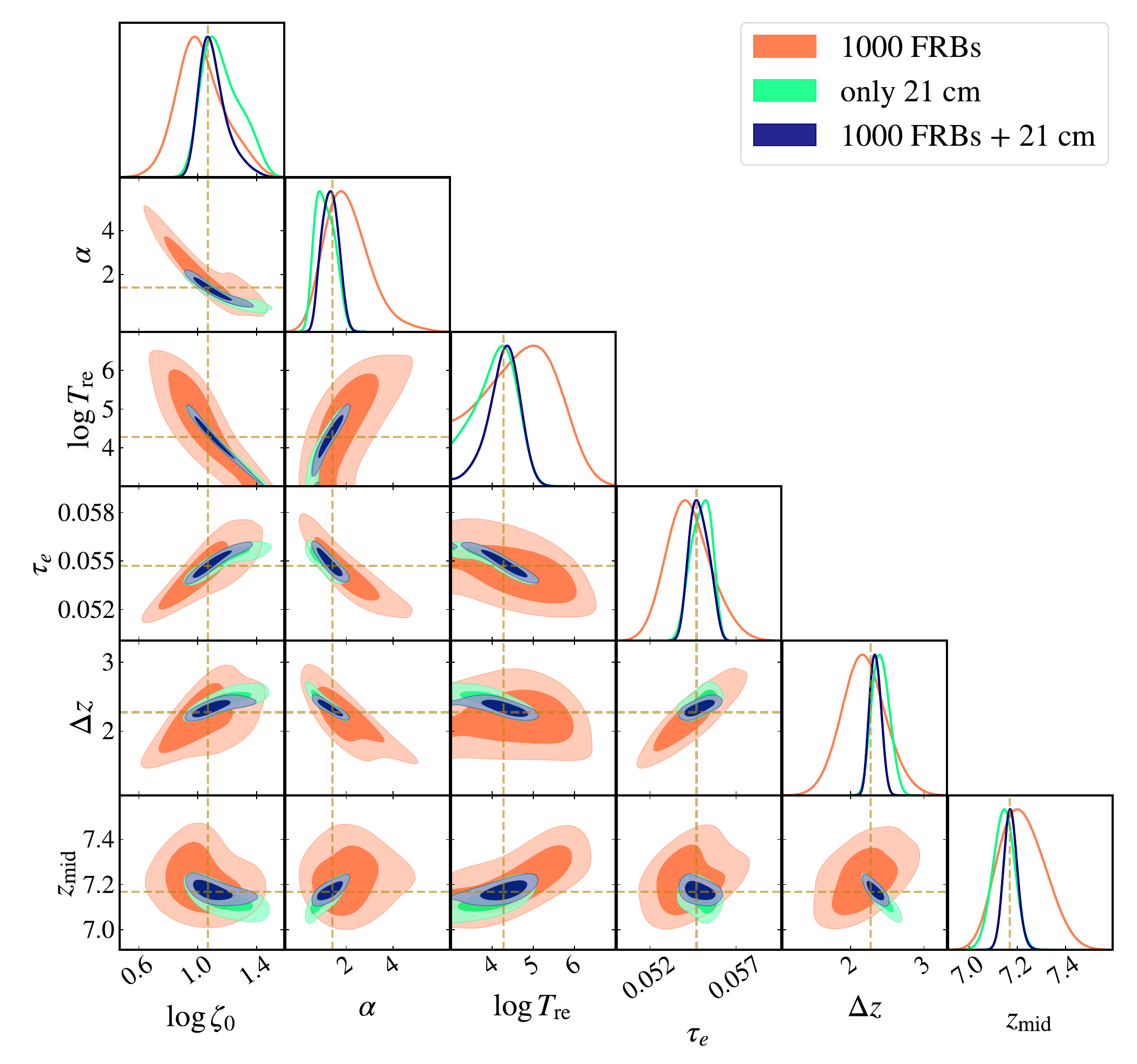}
    \caption{The comparison of parameter posteriors for the cases with 1000 FRBs (same as Figure \ref{fig:corner_constr}), only 21 cm, and joint analysis with 1000 FRBs + 21 cm.  The rest of the details are the same as Figure \ref{fig:corner_constr} except the input parameters have been shown in \textit{golden} dashed line for clarity. }
    \label{fig:corner_constr_21}
\end{figure*}
\begin{figure*}
    \centering
    \includegraphics[width=0.85\textwidth]{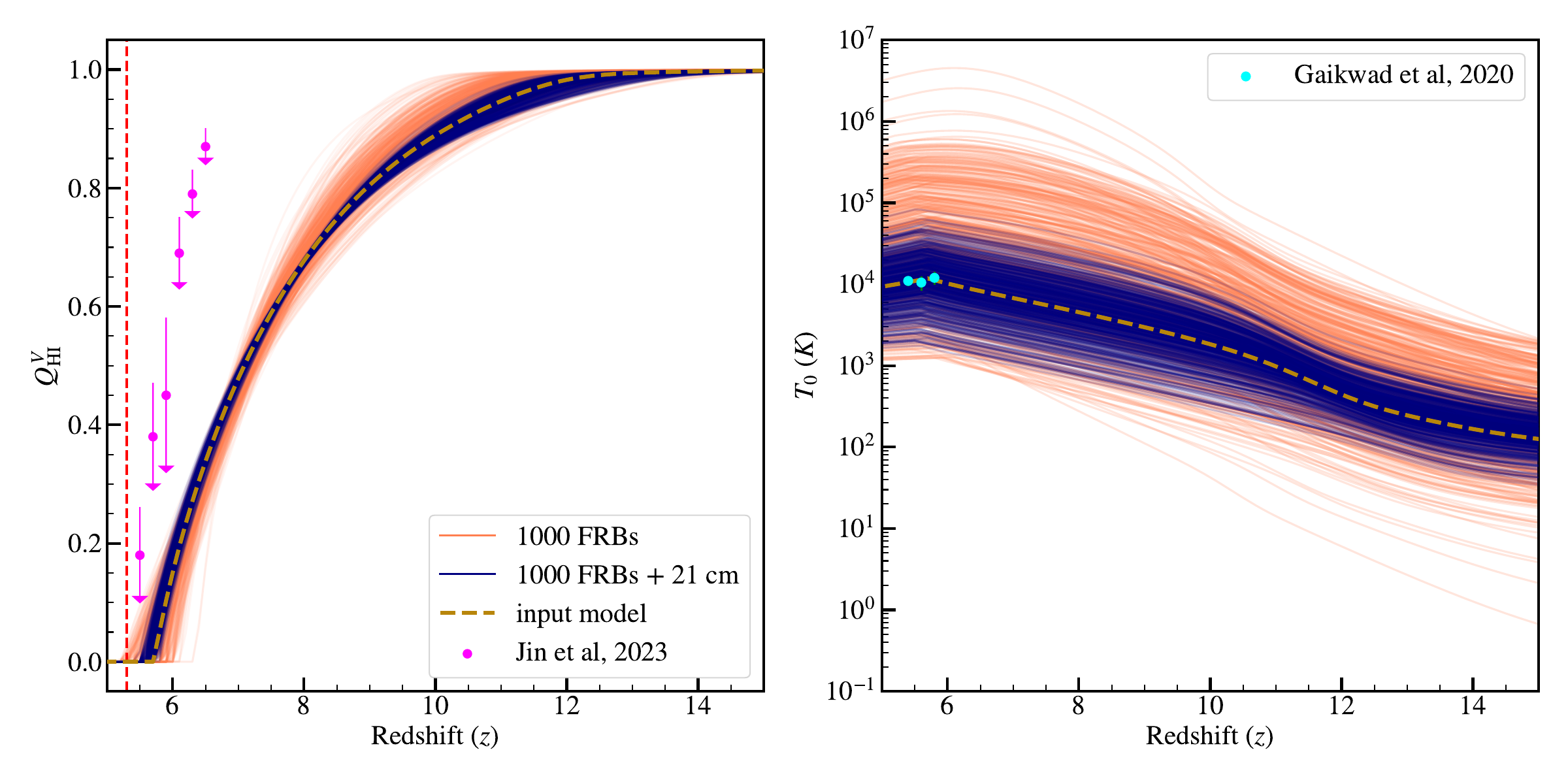}
    \caption{The posterior model comparison plots for the 1000 FRBs mocks (\textit{orange}) and 1000 FRBs + 21 cm (\textit{blue}). The input models are shown in \textit{golden} dashed lines. The rest of the descriptions are the same as Figure \ref{fig:posterior_models} above.}
    \label{fig:posterior_models_21cm}
\end{figure*}
\subsection{Constraining parameters with mock FRBs}
\label{sec:params_cons}
\begin{table*}
\centering
\renewcommand{\arraystretch}{0.5}
\setlength{\tabcolsep}{6pt}
\begin{threeparttable}
\begin{tabular}{ccccccccc}
\hline
Cases & Parameters & Prior & Input & Mean & 68\% Limits& 95\% Limits& Best-fit\\
\hline  
\multirow{16}{4em}{100 FRBs}&  & &&&&& \\ 
 &  &  & &&&&& \\ 
 & $\log(\zeta_0)$     & [$0,\infty$]     & $1.079$ & $0.86$ & [$0.57,1.04$]&[$0.45,1.35$]  & 1.07\\ \\
 &$\alpha$ & [$-20,20$]  & $1.4$ & $3.53$ & [$1.73,5.52$] & [$0.19,6.74$] & 1.39 \\ \\
 &$\log T_{\mathrm{re}}$(in K)    & [$3, 7$]      & $4.279$ & $4.77$& [$3.43,5.70$]  & [$3,7$] &4.42 \\ \\
\cline{2-8} \\
 &$\tau_e$& &$0.0547$ & $0.0530$ & [$0.0505,0.0546$]& [$0.0491,0.0580$] & 0.0547\\ \\
 &$\Delta z$& &$2.275$ & $1.79$ & [$1.19,2.14$]&[$0.93,2.90$] & 2.33\\ \\
&$z_{\mathrm{mid}}$ &  &$7.168$ & $7.19$ & [$7.00,7.34$]&[$6.88,7.54$]& 7.18\\ \\
\hline \\
\hline \\
\multirow{15}{4em}{500 FRBs}&  & &&&&& \\
 & $\log(\zeta_0)$     & [$0,\infty$]     & $1.079$ & $0.96$ & [$0.76,1.16$]&[$0.55,1.33$] & 1.20 \\ \\
 &$\alpha$ & [$-20,20$]  & $1.4$ & $2.6$ & [$0.91,3.67$]  & [$0.15,5.65$] & 1.06 \\ \\
 &$\log T_{\mathrm{re}}$(in K)    & [$3, 7$]      & $4.279$ & $4.75$ & [$3.78,5.83$] & [$3,6.11$] &3.78 \\ \\
 \cline{2-8} \\
 &$\tau_e$& &$0.0547$ & $0.0537$ & [$0.0518,0.0551$]& [$0.0506,0.0571$] & 0.0553\\ \\
&$\Delta z$& &$2.275$ & $2.03$ & [$1.56,2.41$]&[$1.27,2.87$] & 2.29\\ \\
&$z_{\mathrm{mid}}$ &  &$7.168$ & $7.20$ & [$7.08,7.31$]&[$7.00,7.43$]&7.16 \\ \\
 \hline \\
 \hline \\
 \multirow{15}{4em}{1000 FRBs}&   \\
 & $\log(\zeta_0)$     & [$0,\infty$]     & $1.079$ & $1.02$ & [$0.84,1.16$] & [$0.73,1.38$] & 1.07\\ \\
 &$\alpha$ & [$-20,20$]  & $1.4$ & $2.04$ & [$0.95,2.80$]  & [$0.27,4.06$] & 1.44\\ \\
 &$\log T_{\mathrm{re}}$(in K)    & [$3, 7$]      & $4.279$ & $4.69$ & [$3.85,5.69$] & [$3,5.92$] & 4.32 \\ \\
 \cline{2-8} \\
 &$\tau_e$& &$0.0547$ & $0.0543$ & [$0.0528,0.0554$]& [$0.0518,0.0569$] & 0.0547\\ \\
&$\Delta z$& &$2.275$ & $2.18$ & [$1.87,2.46$]&[$1.60,2.79$] &  2.27\\ \\
&$z_{\mathrm{mid}}$ &  &$7.168$ & $7.22$ & [$7.11,7.31$]&[$7.04,7.41$]& 7.17\\ \\

\hline \\ 
 \hline \\
 \multirow{15}{6em}{only 21 cm }&   \\
 & $\log(\zeta_0)$     & [$0,\infty$]     & $1.079$ & $1.16$ & [$1.00,1.26$] & [$0.95,1.43$] & 1.13\\ \\
 &$\alpha$ & [$-20,20$]  & $1.4$ & $1.09$ & [$0.60,1.46$]  & [$0.38,1.89$] & 1.10\\ \\
 &$\log T_{\mathrm{re}}$(in K)    & [$3, 7$]      & $4.279$ & $4.07$ & [$3.64,4.67$] & [$3,4.84$] & 4.17 \\ \\
 \cline{2-8} \\
 &$\tau_e$& &$0.0547$ & $0.0550$ & [$0.0544,0.0557$]& [$0.0538,0.0560$] & 0.0550\\ \\
&$\Delta z$& &$2.275$ & $2.41$ & [$2.28,2.52$]&[$2.20,2.64$] &  2.39\\ \\
&$z_{\mathrm{mid}}$ &  &$7.168$ & $7.14$ & [$7.10,7.19$]&[$7.05,7.23$]& 7.15\\ \\
\hline \\ 
 \hline \\
 \multirow{15}{6em}{1000 FRBs + 21 cm}&   \\
 & $\log(\zeta_0)$     & [$0,\infty$]     & $1.079$ & $1.10$ & [$0.99,1.17$] & [$0.94,1.32$] & 1.09\\ \\
 &$\alpha$ & [$-20,20$]  & $1.4$ & $1.31$ & [$0.92,1.68$]  & [$0.65,1.97$] & 1.30\\ \\
 &$\log T_{\mathrm{re}}$(in K)    & [$3, 7$]      & $4.279$ & $4.28$ & [$3.96,4.72$] & [$3.42,5.00$] & 4.30 \\ \\
 \cline{2-8} \\
 &$\tau_e$& &$0.0547$ & $0.0549$ & [$0.0542,0.0554$]& [$0.0538,0.0559$] & 0.0548\\ \\
&$\Delta z$& &$2.275$ & $2.34$ & [$2.25,2.41$]&[$2.19,2.49$] &  2.33\\ \\
&$z_{\mathrm{mid}}$ &  &$7.168$ & $7.17$ & [$7.14,7.20$]&[$7.12,7.23$]& 7.17\\ \\
 \hline \\
\end{tabular}
\end{threeparttable}
\caption{Parameter constraints obtained from the MCMC-based analysis for different scenarios. For each case, the first three rows correspond to the free parameters of the model while the rest three are the derived parameters. The free parameters are assumed to have uniform priors with the ranges mentioned in the third column. The other respective columns show the input value for mock generation and the recovered posterior mean value along with 68\% and 95\% confidence limits on different parameters. The best fit values are quoted in the last column.}
\label{tab:param_cons}
\end{table*}

In this section, we discuss our results for the forecast on reionization parameter constraints using the FRB mock dataset. We study three cases with different numbers of FRB samples i.e. 100, 500, and 1000. We generate these samples assuming a reionization history corresponding to a fiducial set of model parameters as mentioned above in Section \ref{sec:generate_mock}. For this analysis, we keep the clumping factor to be fixed at $C_{\mathrm{HII}}=3$. We choose a wide range of priors as specified in Table \ref{tab:param_cons}.

In Figure \ref{fig:corner_constr}, we show the recovered parameter posteriors from FRB mocks for the three different numbers of samples (100 : \textit{black}, 500 : \textit{green} \& 1000 : \textit{orange}). Each of these includes three free parameters ($\log \zeta_0$, $\alpha$ and $\log T_{\mathrm{re}}$) and three derived parameters (i.e. $\tau_e$, $\Delta z$ and $z_{\mathrm{mid}}$). We find that most of the input parameters are reasonably well recovered by the posteriors (see the 6th and 7th columns in Table \ref{tab:param_cons} for 68\% and 95\% confidence intervals) using the FRB mocks. We check that even only 100 samples can constrain the ionizing efficiency parameters well within 95\% confidence limits (i.e. $\log \zeta_0=0.86^{+0.49}_{-0.41}$ and $\alpha=3.53^{+3.21}_{-3.33}$). On the other hand, the recovered posterior on temperature increment parameter ($T_{\mathrm{re}}$) is constrained from both sides at 68\% confidence, although it doesn't provide any constraint at 95\% confidence. This behaviour can be explained by the fact that DM evolution doesn't vary much with respect to $T_{\mathrm{re}}$  as  we have discussed earlier (in Section \ref{sec:DM_variation}). Yet, some of the extreme $T_{\mathrm{re}}$ models are discarded by the priors we assume on reionization start and end. However, the derived timeline of reionization is well consistent within the uncertainties ($\Delta z=1.79^{+1.11}_{-0.86}$, $z_{\mathrm{mid}}=7.19^{+0.35}_{-0.31}$ and $\tau_e = 0.0530^{+0.005}_{-0.0039}$ at 95\% confidence). As we increase the number of FRB samples (i.e. 500 and 1000), all the posterior constraints improve slightly along with providing upper limits on $\log T_{\mathrm{re}}$ parameter ($\log T_{\mathrm{re}}$ at 95\% confidence is $\le 6.11$ and $\le 5.92$ for 500 and 1000 FRBs respectively). A higher number of mock FRB samples also discard some of the models with extreme evolutions of ionizing efficiency. These further tighten up the recovered posteriors on the ionizing efficiency parameters (i.e. $\log \zeta_0=1.02^{+0.36}_{-0.30}$ and $\alpha=2.04^{+2.02}_{-1.77}$ at 95\% confidence for 1000 samples). Similarly, the uncertainties on the derived posteriors for duration and midpoint of reionization are reduced, providing $\Delta z=2.18^{+0.60}_{-0.58}$ and $z_{\mathrm{mid}}=7.22^{+0.19}_{-0.18}$ at 95\% confidence with 1000 samples. The uncertainties on reionization interval even reach $\lesssim 15\%$  at a credibility of 68\% ($\Delta z=2.18^{+0.28}_{-0.31}$). These numbers are in a similar ballpark as quoted in earlier study \citep{2021MNRAS.505.2195P} with $10^4-10^5$ FRB samples and different inhomogeneous reionization models. The constraints on CMB scattering optical depth become stringent as well ($\tau_e = 0.0543^{+0.0026}_{-0.0022}$) which is comparable with the expectations from upcoming CMB experiments. 

These results highlight the power of high redshift FRBs in constraining the sources and nature of the reionization. These are more evident from the plots in Figure \ref{fig:posterior_models}, where we have shown 500 random models from the posterior chains for the two cases (using 100 and 1000 FRB mocks for clarity). From the \textit{left} panel, it is evident that some of the models with very fast evolution and relatively early reionization end are allowed with 100 FRB mocks (shown in \textit{black}) although these are not favored once we increase the mock samples (shown in \textit{orange}). However, all the allowed models are well within the recent nearly model independent limits on IGM neutral fraction using Ly-$\alpha$ and Ly-$\beta$ forests by \citep{2023ApJ...942...59J} even if we do not use these limits in our MCMC runs. From the \textit{right} panel, we can observe that the constraints on the IGM mean temperature ($T_0$) evolution do not show significant improvement which is a consequence of the fact that $T_{\mathrm{re}}$ is not well constrained by the FRB data as discussed before. On the contrary, the already available constraints on the mean IGM temperature from Ly-$\alpha$ spike statistics study around redshift $z\sim 5-6$ \citep{2020MNRAS.494.5091G} are much tighter which has the potential to be a complementary probe of IGM thermal evolution \citep{2022MNRAS.515..617M}.

\subsection{Forecasting with 21~cm + FRB mocks }
Now, we move further and study the prospects for the inclusion of large scale 21~cm information in the likelihood analysis. We perform the analysis using `only 21 cm' data followed by the joint prospects of `1000 FRBs + 21 cm'. 

In Figure \ref{fig:corner_constr_21}, we show the recovered parameter posteriors for these two cases (i.e. `only 21 cm' in \textit{lime green} and  `1000 FRBs + 21 cm' in  \textit{blue}) on top of the earlier posterior distribution with only 1000 FRBs (\textit{orange}). We quantify the parameter recoveries along with 68\% and 95\% uncertainties at the final two rows in Table \ref{tab:param_cons}. We find that 21~cm power spectra at three redshifts alone can provide improved constraints on the astrophysical parameters as well as the reionization timeline. The 2D posterior comparison of $\log \zeta_0-\alpha$ clearly shows that the reionization models with relatively fast evolution (i.e. high $\alpha$ and smaller $\zeta_0$) are not preferred by the 21 cm mock data. This increases the strength of the constraints on the marginalised 1D posterior distribution for the above two parameters (i.e.  $\log \zeta_0=1.16^{+0.27}_{-0.21}$ and $\alpha=1.89^{+0.80}_{-0.71}$ at 95\% significance). This is further evident from the posterior distribution of the reionization interval ($\Delta z=2.41^{+0.23}_{-0.21}$) where the lower values ($\Delta z \lesssim 2.0$) allowed by the FRB data, are discarded once the 21 cm data are included. As a further consequence, this also improves the constraints on the reionization temperature increment parameter ($\log T_{\mathrm{re}}\le 4.84$) by disfavoring the extremely high temperature models. This is expected as the temperature evolution maintains the recombination rate at each redshift which can further control the amplitude of 21 cm fluctuations.  The large scale power spectra also significantly alleviate the degeneracy between CMB scattering optical depth ($\tau_e= 0.0550^{+0.0010}_{-0.0012}$) and reionization midpoint ($z_{\mathrm{mid}}=7.14^{+0.09}_{-0.07}$). Further, we note that the directions of degeneracy between $\Delta z$ and $z_{\mathrm{mid}}$ are opposite for FRB DM mocks and 21 cm mocks. Hence, this degeneracy is broken when we include both the data sets in likelihood for the joint analysis (i.e. `1000 FRBs + 21 cm' case). The joint posterior distribution further shrinks the parameter spaces providing stringent constraints on astrophysical parameters and reionization timeline. Specifically, extremely low  $T_{\mathrm{re}}$  models are now disfavored by the joint posteriors placing a constraint ($3.42\le\log T_{\mathrm{re}}\le5.00$) within the specified prior range even at the 95\% confidence level. The improvements are slight in case of the ionizing efficiency parameter in comparison to the `only 21 cm' case, providing  $\log \zeta_0=1.10^{+0.22}_{-0.16}$ and $\alpha=1.31^{+0.66}_{-0.66}$ at 95\% significance. Not surprisingly, the recovered posteriors for the reionization parameter are constrained with a precision of $\lesssim 8\%$ ($\Delta z=2.34^{+0.15}_{-0.14}$ \& $z_{\mathrm{mid}}=7.17^{+0.06}_{-0.05}$).

The above discussions can be more evident from the plots shown in Figure \ref{fig:posterior_models_21cm}, where we show the ionization and thermal history of the 500 random models from the recovered posteriors. From the left panel, it is visually apparent that the uncertainties on reionization histories are significantly improved for the samples from the joint posterior (in \textit{blue}). The same is true for the temperature evolution which is shown in the right panel. The allowed range is still much wider than the uncertainties provided by the observational study \citep{2020MNRAS.494.5091G}. However, this is not very surprising as we have used a conservative uncertainty on the 21 cm mocks ($10\%$ of the fiducial value) along with considering only a single large scale mode expected to be available from the future observations.

Lastly, it is important to highlight that we have fixed the strength of recombination by keeping clumping factor $C_{\mathrm{HII}}$ fixed throughout the analysis. This parameter is highly degenerate with the ionizing efficiency parameter ($\zeta$) which can weaken the derived constraints on the different parameters.  Although 21 cm fluctuation information is expected to alleviate some of these degeneracies, it is ideal to use all the available $k$ modes including the small scale information while doing the parameter space exploration. However, for this proof concept study, we don't vary this parameter and plan to pursue a separate study with more observables in the future. Also, as mentioned earlier in Section \ref{sec:params_explore}, we use two observationally motivated priors on the neutral fraction evolution of the IGM (reionization start and end) which contribute to discarding some of the extreme reionization models. 
\section{Summary \& Conclusions}
\label{sec:conc}
The Epoch of Reionization is one of the final missing pieces in the evolutionary history of our universe. There are a handful of observational probes across multiple wavelength regimes which have been proven to be extremely useful to get insights into this high redshift era. In the near future, with the advancement of cutting-edge observational facilities, we expect to pin down the mysteries of the reionization era including the exact timeline. The highly dispersed FRBs and the 21~cm power spectra from the fluctuations in neutral hydrogen atom field are the two potential probes which can shed light on the reionization in the upcoming days. In this work, we study the prospects of these two probes in constraining the thermal and ionization history of the universe during EoR. Below, we pointwise summarize our main findings.

\begin{itemize}
    \item We utilize a realistic semi numerical reionization model based on a photon conserving algorithm (containing the inhomogeneities due to recombination and radiative feedback effects) to study the expected dispersion measure variation for high redshift FRBs ($z\ge 7$).  We check that the different reionization model parameters can give rise to different DM evolution which can be further used to put constraints on reionization models. These findings are similar to the earlier studies in the literature with excursion set based semi numerical reionization model \citep{2021MNRAS.505.2195P}.
    
    \item Next, we generate mock FRBs at high redshifts ($7\le z\le 15$) and simulate their DM uncertainties assuming the realistic expectation for the upcoming telescopes following the layout provided in \citet{2022ApJ...933...57H}. We pursue a forecasting study with 100, 500, and 1000 FRB samples to gauge the viability of FRBs as a reionization probe. We find that the posterior with 100 FRBs can recover the input model parameters (except the thermal parameter, $T_{\mathrm{re}}$ at 95\% confidence) along with the derived parameters quantifying reionization timeline. With 1000 FRBs, these constraints improve significantly with $\lesssim 30\%$ uncertainty on the reionization interval ($\Delta z$) and percentage level uncertainty on the reionization midpoint ($z_{\mathrm{mid}}$) while providing an upper limit on $T_{\mathrm{re}}$ which helps to discard some extreme temperature evolution models.
    
    \item Lastly, we study the effects after the inclusion of large scale ($k\sim~0.14~h/\mathrm{cMpc}$) 21 cm power spectra on the parameter constraints. As a realistic (slightly conservative) approach, we assume 10\% uncertainties on the power spectra at three different redshifts (at $z\sim7,~8~ \&~ 9$). We show that 21 cm power spectra alone can rule out significant fraction of parameter spaces which are allowed by DM likelihood of 1000 FRB mocks. The remaining degeneracies can be alleviated by the joint likelihood of 21 cm and FRB DMs. We find that the joint posterior can provide constraints on the thermal parameter from both ends even at a 95\% significance level. This further brings down the recovered $\Delta z$ uncertainties to $\lesssim 8\%$ at 95\% confidence level.
\end{itemize}
According to the present consensus, it is really uncertain to get an accurate estimate of the detectable high redshift FRBs. If these objects indeed exist in the high redshift universe, they can be detected with SKA  at a rate of 100/day/sky assuming standard cosmic star formation rate density \citep{2020MNRAS.497.4107H} for redshift $z\gtrsim6$. While the deployment work for SKA-low has already been started, it is expected to take around a few more years to be fully operational. A similar timeline has been set for other large facilities like Giant Magellan Telescope (GMT), and Extremely Large Telescope (ELT) which can be critical to determine the redshifts of the sources. In this context, our
 study provides a proof of concept for using high redshift FRBs and large scale 21 cm power spectra as simultaneous probes of the reionization epoch. These show the potential to constrain both the ionization and thermal history (albeit weakly) of the universe during hydrogen reionization. However, the exact precision of these constraints is subject to change depending upon the true measurements of DM uncertainties by future observations (i.e.  larger uncertainties on DM values will weaken the constraints on the parameters and vice versa).
 
 As a next step, we would like to extend our analysis including the helium reionization epoch ($2\lesssim z\lesssim 4$) which is expected to be accessible soon by the discoveries of new FRBs with advanced facilities. However, that will require detailed modelling of the IGM including much complex astrophysics. On the other hand, we also target to utilize the 21 cm information in its full glory including small scales along with FRB clustering. This will allow us to free other astrophysical parameters providing a more realistic reionization scenario. Nevertheless, this requires more efficient models/techniques to pursue parameter exploration utilizing high resolution simulation which sets the future direction of our research. 


\section*{Data Availability}

The data presented in this article will be shared on reasonable request to the corresponding author (BM).
\begin{acknowledgements}
     The author (BM) thanks Prof. Tirthankar Roy Choudhury for the guidance with reionization modelling
and Dr. Apurba Bera for the useful discussions on the aspects of FRBs.
BM acknowledges the support of the Max Planck Group.
\end{acknowledgements}

-------------------------------------------------------------------
\bibliographystyle{aa}
\bibliography{frb_aanda}
\end{document}